\documentclass[a4paper,superscriptaddress,notitlepage]{revtex4-2}

\usepackage[pdftex]{graphicx}
\usepackage{textcomp} 
\usepackage{dsfont}
\usepackage[a4paper, right=4cm, left=4cm, top=2.5cm, bottom=2.5cm]{geometry}
\usepackage[usenames]{color}
\usepackage{subcaption}

\definecolor{RED}{rgb}{0.7,0.0,0.0}
\definecolor{GREEN}{rgb}{0,0.4,0}
\def\comm#1{{#1}}
\def\commSDI#1{{#1}}
\def\commSDIforarxiv#1{{\color{GREEN}#1}}

\def\L{L_\mathrm{g}} 

\newcommand{\LOA}{Laboratoire d'Optique Appliqu\'ee, ENSTA Paris, CNRS, Ecole Polytechnique, Institut Polytechnique de Paris, 181 chemin de la Huni\`ere et des Joncherettes, 91120 Palaiseau, France}
\newcommand{\Ardop}{Ardop Engineering, Cit\'e de la Photonique, 11 Avenue de la Canteranne, b\^at. Pl\'eione, 33600 Pessac, France}
\newcommand{\WIS}{Department of Physics of Complex Systems, Weizmann Institute of Science, Rehovot 7610001, Israel}

\begin{document}

\title{High-Harmonic Generation and Correlated Electron Emission from Relativistic Plasma Mirrors at 1~kHz Repetition Rate}

\author{S. Haessler}\email{stefan.haessler@cnrs.fr}\affiliation{\LOA} 
\author{M. Ouill\'e}\affiliation{\LOA}\affiliation{\Ardop}
\author{J. Kaur}\affiliation{\LOA}
\author{M. Bocoum}\affiliation{\LOA}
\author{F. B\"ohle}\affiliation{\LOA}
\author{D. Levy}\affiliation{\WIS}
\author{L. Daniault}\affiliation{\LOA}
\author{A. Vernier}\affiliation{\LOA}
\author{J. Faure}\affiliation{\LOA}
\author{R. Lopez-Martens}\affiliation{\LOA}

\begin{abstract}
We report evidence for the first generation of XUV spectra from relativistic surface high-harmonic generation (SHHG) on plasma mirrors at a kilohertz repetition rate, emitted simultaneously with energetic electrons. SHHG spectra and electron angular distributions are measured as a function of the experimentally controlled plasma density gradient scale length $\L$ for three increasingly short and intense driving pulses: 24~fs and $a_0=1.1$, 8~fs and $a_0=1.6$, and finally 4~fs and $a_0\approx2.1$, \comm{where $a_0$ is the peak vector potential normalized by $m_\mathrm{e} c/e$ with the elementary charge $e$, the electron rest mass  $m_\mathrm{e}$, and the vacuum light velocity $c$}. For all driver pulses, we observe correlated relativistic SHHG and electron emission in the range \commSDI{$\L\in[\lambda/25,\lambda/5]$}, with an optimum gradient scale length of \commSDI{$\L\approx\lambda/10$}. 
\end{abstract}

\maketitle
\section{Introduction}
\comm{Surface high-harmonic generation (SHHG) from relativistic plasma mirrors~\cite{dromey_high_2006,tarasevitch_transition_2007,teubner_high-order_2009,thaury_high-order_2010} is a promising method for greatly enhancing the power of attosecond light pulses. This is motivated by key properties that differentiate it from the established method based on high-harmonic generation in gases~\cite{calegari_advances_2016}. Since the medium is already highly ionized, SHHG has no inherent limitation for the driving intensity such that a large fraction of the energy of an ultra-high intensity driving laser can be compressed into an attosecond electromagntic pulse. This is expected to occur with extremely high, percent-level conversion efficiencies~\cite{tsakiris_route_2006, anderBruegge2010nanobunching,chopineau_identification_2019,edwards_x-ray_2020} when driven under strongly relativistic conditions with a normalized vector potential $a_0 = e A_0/ m_\mathrm{e} c \gg1$ [a practical formula is $a_0=\sqrt{I [\mathrm{W cm}^{-2}] \,\lambda^2 [\mu \mathrm{m}^2]/(1.37\times10^{18})}$], where $e$ is the elementary charge,  $m_\mathrm{e}$ the electron rest mass, $c$ the vacuum light velocity, $A_0$ the laser's peak vector potential, $I$ its peak intensity and $\lambda$ its central wavelength. Reported experimental laser-to-XUV conversion efficiencies for plasma mirrors with $a_0\gtrsim1$ are $\sim10^{-4}$~\cite{Roedel2012ultrasteep,Heissler2014multimuJ,yeung_experimental_2017,jahn_towards_2019} for photon energies $>20\:$eV, but are expected to increase with higher-intensity drivers. Finally, relativistic SHHG produces intrinsically Fourier-limited attosecond-pulses, such that the complete spectral bandwidth, from the spectral cutoff in the XUV (or even hard x-rays~\cite{dromey_bright_2007}) all the way down to lowest harmonics and even the fundamental laser frequency, can contribute to the formation of attosecond pulses without costly spectral filtering and compression~\cite{gordienko_relativistic_2004,bohle_generation_2020,chopineau_spatio-temporal_2021}.}

Reaching the relativistic SHHG regime with $a_0>1$ requires an on-target intensity of $\gtrsim10^{18}\:$W/cm$^2$ for an 800-nm laser while retaining a very steep surface plasma density profile
\begin{equation}
     n(x) = n_\mathrm{c} \exp\left[ x/\L \right],
\label{eq:densityprofile}
\end{equation}
with a scale length $\L$ of a small fraction of the driving laser wavelength $\lambda$. Here $n_c$ is the nonrelativistic critical plasma density for the driving wavelength and $x$ is the coordinate in the target normal direction. Technically this requires a highly focusable terawatt-class laser with a temporal contrast of $\gtrsim10^{10}$. These conditions are typically met by Joule-class amplifier chains with dedicated contrast filters~\cite{kapteyn_prepulse_1991,dromey_plasma_2004,thaury_plasma_2007,jullien_xpw_2005} and operating at $\lesssim10\:$Hz repetition rate~\cite{thaury_plasma_2007,Roedel2012ultrasteep,dollar_scaling_2013,Heissler2014multimuJ,yeung_experimental_2017,jahn_towards_2019,chopineau_identification_2019}.

Many applications as well as parametric studies of this regime would benefit from a higher repetition rate. At LOA, we have developed a unique laser chain with a power-scaled hollow-core-fiber postcompression system~\cite{bohle_compression_2014,ouille_relativistic-intensity_2020} operating at 1~kHz repetition rate. Using this kHz-laser, which achieves ultra-high intensities with few-mJ pulse energy and few-cycle pulse duration, we have demonstrated laser-plasma interaction in the relativistic regime through laser-wakefield acceleration of electrons, both in underdense gas jets~\cite{guenot_relativistic_2017,gustas_high-charge_2018} and in the underdense part of a smooth plasma density gradient on a plasma mirror~\cite{zaim_few-cycle_2019}. Here, we report on the first experimental demonstration of relativistic SHHG at kHz-repetition rate, the arguably most demanding application in terms of laser performance, as it depends critically on the spatio-temporal pulse quality and the temporal contrast.

\subsection{High-harmonic generation on plasma mirrors} 
For relativistic driving intensities, $a_0\gtrsim1$, the SHHG emission mechanism is described by a push--pull process~\cite{,thevenet_physics_2016}, also dubbed ``relativistic electron spring''~\cite{gonoskov_ultrarelativistic_2011,gonoskov_theory_2018}, repeating once per driving laser period. The laser field first pushes electrons into the plasma, piling up a dense electron bunch and creating a restoring internal plasma field. As the laser field changes sign, the combined plasma and laser fields accelerate the electron bunch to a relativistic velocity towards the vacuum. SHHG emission then either results from a pure phase modulation of the incident laser field by the relativistically moving critical-density plasma surface [``relativistic oscillating mirror'' (ROM)]~\cite{lichters_short-pulse_1996,gordienko_relativistic_2004,tsakiris_route_2006}, or as a coherent synchrotron emission (CSE) from the dense relativistic electron bunch as it gets accelerated orthogonally by the laser electric field~\cite{anderBruegge2010nanobunching, mikhailova_isolated_2012,edwards_x-ray_2020}.

For $a_0\sim1$, this relativistic SHHG emission needs to be discriminated from coherent wake emission (CWE)~\cite{quere_coherent_2006,thaury_high-order_2010}, generated already at sub-relativistic intensities, $a_0<1$. Once per laser field cycle, Brunel electrons~\cite{brunel_not-so-resonant_1987} are accelerated out and back into the surface plasma, to form via collective trajectory crossings an electron density peak traversing the overdense part ($n(x)>n_c$) of the plasma density gradient. In its wake, it excites plasma oscillations at the local plasma frequency which finally lead to the emission of one attosecond light pulse per laser cycle.  Clear criteria are known for the discrimination of the two types of SHHG emission.

\emph{(i) Intensity dependence:} The laser-to-SHHG conversion efficiency of CWE is constant over a large intensity range $>10^{15}\:$W/cm$^2$~\cite{thaury_high-order_2010}, whereas that of relativistic SHHG increases rapidly over a large $a_0$-range from $a_0\sim1$ until a saturation in the the ultra-relativistic limit $a_0\sim100$~\cite{thaury_high-order_2010,edwards_x-ray_2020}. Note however that in this experimentally still out-of-reach ultra-relativistic regime, the radiation reaction force may significantly modify SHHG~\cite{tang_plasma_2017}.

\emph{(ii) Phase properties:} The intensity-dependence of the Brunel trajectories leads to increasing delays between successive attosecond pulses under the driving pulse envelope, and thus to frequency-broadened negatively chirped individual harmonics~\cite{quere_coherent_2006,varju_frequency_2005}. A positively chirped driving pulse can partially compensate the effect and consequently lead to spectrally narrower CWE-harmonics~\cite{quere_coherent_2006,malvache_coherent_2013,mitrofanov_chirp-controlled_2020}.  In contrast, for moderate $a_0\sim1$, the relativistic SHHG emission is Fourier-limited~\cite{thaury_high-order_2010,chopineau_spatio-temporal_2021} such that despite its much steeper intensity dependence and consequently shorter temporal envelope, its harmonic spectral width is narrower than that of CWE harmonics generated in the same conditions~\cite{thaury_plasma_2007,thaury_high-order_2010,kahaly_direct_2013}. Only for higher $a_0\gtrsim10$, \comm{while the individual attosecond pulses remain Fourier-limited, plasma surface denting will again lead to a varying pulse-spacing and thus induce an intrinsic harmonic chirp}, with the opposite sign compared to that of CWE~\cite{thaury_high-order_2010,behmke_controlling_2011,vincenti_optical_2014}.  

\emph{(iii) Dependence on the plasma density gradient:} The density gradient of the plasma-vacuum interface is a most crucial parameter~\cite{zepf_role_1998} and its experimental control is a prerequisite for efficient SHHG~\cite{Roedel2012ultrasteep,dollar_scaling_2013,kahaly_direct_2013}. In CWE, the electron density peak formation and conversion of plasma oscillations into XUV light are optimized for a very steep gradient with scale length $\L\sim\lambda/100$. A growing $\L$ causes the CWE-efficiency to drop rapidly~\cite{thaury_high-order_2010} while the harmonic chirp and thus spectral width of the CWE-harmonic peaks increase~\cite{quere_coherent_2006,malvache_coherent_2013}. For the steep gradients optimizing CWE, relativistic SHHG is supressed by the gyromagnetic effect~\cite{geindre_relativistic_2006} preventing electrons from escaping the plasma. Instead, relativistic SHHG has repeatedly been found in single-shot experiments to be optimized by smoother optimal density gradients~\cite{Roedel2012ultrasteep,dollar_scaling_2013,kahaly_direct_2013,chopineau_identification_2019,gao_double_2019}, characterized by scale lengths $\L\approx\lambda/10$. For further increased $\L\gtrsim\lambda/4$, SHHG disappears due to the onset of chaotic electron dynamics~\cite{dollar_scaling_2013,chopineau_identification_2019}.

\emph{(iv) Simultaneous electron emission:} Plasma mirrors also emit electron beams near the specular direction but slightly deviated toward the target normal~\cite{tian_electron_2012,zhou_direct_2021}. The same optimal $\L\approx\lambda/10$ has been found to maximize the charge of relativistic electron bunches emitted from plasma mirrors~\cite{thevenet_vacuum_2016}, which is thus found to be correlated with relativistic SHHG~\cite{thevenet_physics_2016,chopineau_identification_2019} and anti-correlated with CWE~\cite{bocoum_anticorrelated_2016}.

\section{Materials and Methods}
\begin{figure}[tb]
	\centering
	\includegraphics[width=.75\linewidth]{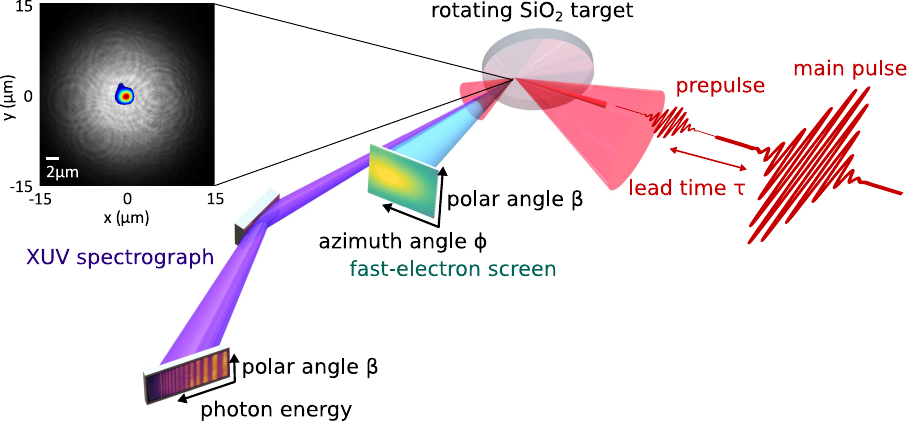}
	\caption{Schematics of the experimental setup for SHHG and electron acceleration on a kHz plasma mirror. The inset shows the on-target intensity profiles of the pre-pulse (in greyscale) and the main-pulse (in colorscale).}
	\label{fig:setup}
\end{figure}

The experiments were carried out using the 1-kHz “Salle Noire” laser system at LOA, delivering pulses with durations adjustable between $24\:$fs and $<4\:$fs and a temporal contrast $>10^{10}$ at $10\:$ps~\cite{ouille_relativistic-intensity_2020}. As illustrated in figure~\ref{fig:setup}, the p-polarized pulses are focused down to a $\approx1.8$-\textmu m FWHM spot by an f/1.3, 30$^\circ$ off-axis parabola onto a rotating optically flat fused-silica target at an incidence of $\theta=55^\circ$. With the pulse energy of $2.6\:$mJ on target, the measured spatial and temporal profiles yield, in the assumed absence of spatio-temporal couplings, peak intensities of $2.5\times10^{18}\:$W/cm$^2$ ($a_0=1.1$ for $\lambda=800\:$nm), $5.5\times10^{18}\:$W/cm$^2$ ($a_0=1.6$ for $\lambda=800\:$nm), and $1.0\times10^{19}\:$W/cm$^2$ ($a_0=2.1$ for $\lambda=770\:$nm) for 24-fs, 8-fs and 4-fs pulses, respectively.

We thus operate in an intensity regime where both CWE and relativistic SHHG are expected to give significant contributions to the generated harmonic signal. Controlling the plasma density gradient scale length lets us modulate their weights~\cite{kahaly_direct_2013}. To this end, a spatially superposed pre-pulse with lead time $\tau$ is focused to a 13-\textmu m FWHM spot in order to prepare a plasma density gradient scale length $\L= L_0+c_\mathrm{s}\tau$~\cite{kahaly_direct_2013,bocoum_sdi_2015}. The expansion speed $c_\mathrm{s}$ is measured using spatial-domain interferometry (SDI)~\cite{bocoum_sdi_2015}, and $L_0$ is the scale length increase owing to the finite temporal contrast of the main pulse. Our observation of CWE (see below) justifies the assumption that $L_0\lesssim\lambda/50$~\cite{thaury_high-order_2010,kahaly_direct_2013}.

\commSDIforarxiv{Note: Compared to the earlier version of this manuscript, the expansion speed $c_\mathrm{s}$ is now a factor $\approx 2$ higher because we have corrected an error in the analysis of our SDI measurements that lead to an underestimation of  $c_\mathrm{s}$. This leaves all qualitative conclusions intact but increases the values of the optimal gradient scale length range. With these increased experimental values, a slightly different definition of the effective penetration depth of the laser into the plasma gradient (cf. eq. \ref{eq:xbeff}) leads to a better match: before we had directly averaged the depth over the pushing half-cycle, while now we just replace the time-varying vector potential in the instantaneous penetration depth by the half-cycle averaged vector potential.}

The emitted SHHG radiation in the specular direction is recorded using an angle-resolving XUV spectrograph (cf. figure~\ref{fig:setup}), home-built with a gold-coated flat-field grating (600 lines/mm, 85.3$^\circ$ incidence) and a micro-channel-plate (MCP) and phosphor screen detector. The MCP is time-gated for $\approx 250\:$ns synchronously with the laser pulses so as to suppress the detection of longer background plasma emission. The phosphor screen is finally imaged by a charged-coupled-device (CCD) camera to record angle-resolved ($\beta\in[-35,35]\:$mrad) SHHG spectra. This only detects the central part of the SHHG beam, which in our conditions has a FWHM-divergence of $\approx70\:$mrad,  owing in particular to the small source spot. The SHHG-spectra shown in the following are angle-integrated over the full detected range and represent the MCP signal without correction for the spectrometer's spectral response so as to enhance the visibility of the higher harmonic orders.

Simultaneously, we detect the angular emission profile of electrons with energies $>150\:$keV with a lanex-screen~\cite{glinec_absolute_2006} covered by a 15-\textmu m aluminium foil, placed between the specular and the target-normal directions and imaged by a CCD camera (cf. figure~\ref{fig:setup}). The distributions along the azimuthal angle $\phi$ (in the incidence plane) shown in the following are integrated over the polar angle $\beta\in[-15^\circ,15^\circ]$ (perpendicular to the incidence plane).

All shown data are acquired by integrating over 100-ms long bursts of pulses at 1~kHz repetition rate. \comm{The carrier-envelope phase (CEP) of the laser pulses has not been locked in this work. Its rapid random variation during these 100-shot acquisitions thus averages out CEP-effects on the SHHG~\cite{kormin_spectral_2018,jahn_towards_2019,bohle_generation_2020} and electron emission.}

\section{Results}

\begin{figure}[h]
    \centering
    \begin{subfigure}{0.4\textwidth}
        \caption{\label{fig:scan24fsa}\hspace*{\textwidth}}
        \includegraphics[width=0.9\textwidth]{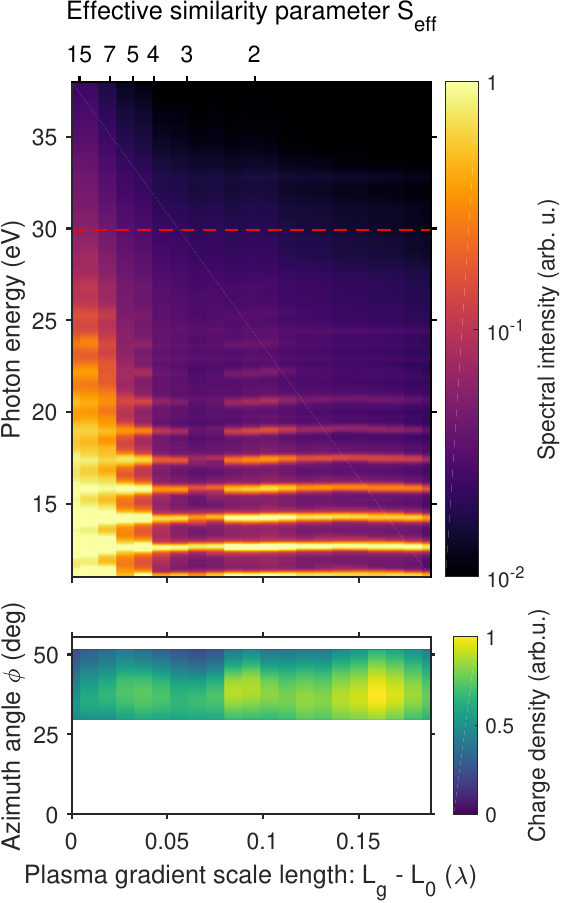}
    \end{subfigure}
    \begin{subfigure}{0.4\textwidth}
        \caption{\label{fig:scan24fsb}\hspace*{\textwidth}}
        \includegraphics[width=0.9\textwidth]{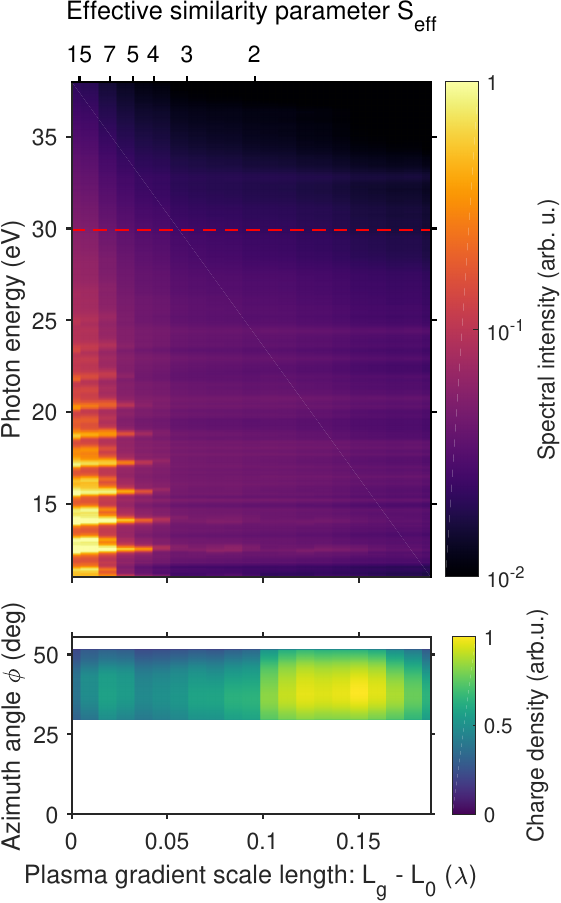}
    \end{subfigure}
    \caption{Synchronously measured SHHG (upper, log-scale) and electron (lower) emission as a function of the plasma gradient scale length. The dashed red line marks the CWE cutoff energy. The normal and specular directions correspond to $\phi=0$ and $\phi=55^\circ$. (\subref{fig:scan24fsa}) For a compressed, 24-fs, $a_0=1.1$ driver pulse, (\subref{fig:scan24fsb}) for the same pulse with added 500~fs$^2$ GDD, 85~fs duration, $a_0=0.6$. The upper horizontal axis gives $S_\mathrm{eff}$ corresponding to $\L$ as given by eq.~\ref{eq:Seff} and assuming $L_0=\lambda/100$.}
    \label{fig:scan24fs}
\end{figure}

Figure~\ref{fig:scan24fs}(a) shows results obtained with a near-Fourier-limited 24-fs driver pulse. The pre-pulse lead time is scanned on the ps-timescale, which translates via an SDI measurement to a change of plasma density gradient scale length from $L_0$ to \commSDI{$L_0+\lambda/5$}, where $\lambda=800\:$nm. Two $\L$-regimes are clearly distinguished. At the shortest gradients, \commSDI{$\L-L_0<\lambda/25$}, the harmonic peaks are broad and their intensity quickly drops with increasing $\L$, as expected for CWE emission. At higher $\L$, the harmonic peaks are remarkably narrower and their intensity goes through an optimum located at \commSDI{$\L-L_0\approx\lambda/10$}, consistent with the theoretical expectations for relativistic SHHG as well as with earlier experimental observations with similar multi-cycle driver pulses~\cite{Roedel2012ultrasteep,kahaly_direct_2013,chopineau_identification_2019,gao_double_2019}. No harmonics are visible beyond the CWE-cutoff at 30~eV, which is not very surprising as the laser intensity is only just around the relativistic threshold, $a_0=1.1$. Finally, we find a stronger electron emission at softer gradients, correlated to the (presumed) relativistic SHHG emission.

We can corroborate this interpretation by using a ``Dazzler'' acousto-optic programmable dispersive filter~\cite{verluise_dazzler_2000} in our amplifier chain to add 500~fs$^2$ GDD to the driving pulses, thus chirping them positively. This stretches them to 85~fs duration and consequently decreases the intensity to $0.7\times10^{18}\:$W/cm$^2$ ($a_0=0.6$ for $\lambda=800\:$nm). As shown in figure~\ref{fig:scan24fs}(b), the harmonic spectral widths decrease significantly at the shortest $\L$, but remain similar in intensity. This is consistent with the compensation of the aperiodicity of the CWE attosecond pulse train and the constant CWE conversion efficiency. The disappearance of the SHHG emission at softer gradients is consistent with the highly nonlinear intensity-dependence of relativistic SHHG. Electrons are still emitted preferentially for the softer gradients and the resulting anti-correlation of harmonics and electrons is reminiscent of our earlier sub-relativistic-intensity results~\cite{bocoum_anticorrelated_2016}.

\begin{figure}[tb]
	\centering
	\includegraphics[width=.5\linewidth]{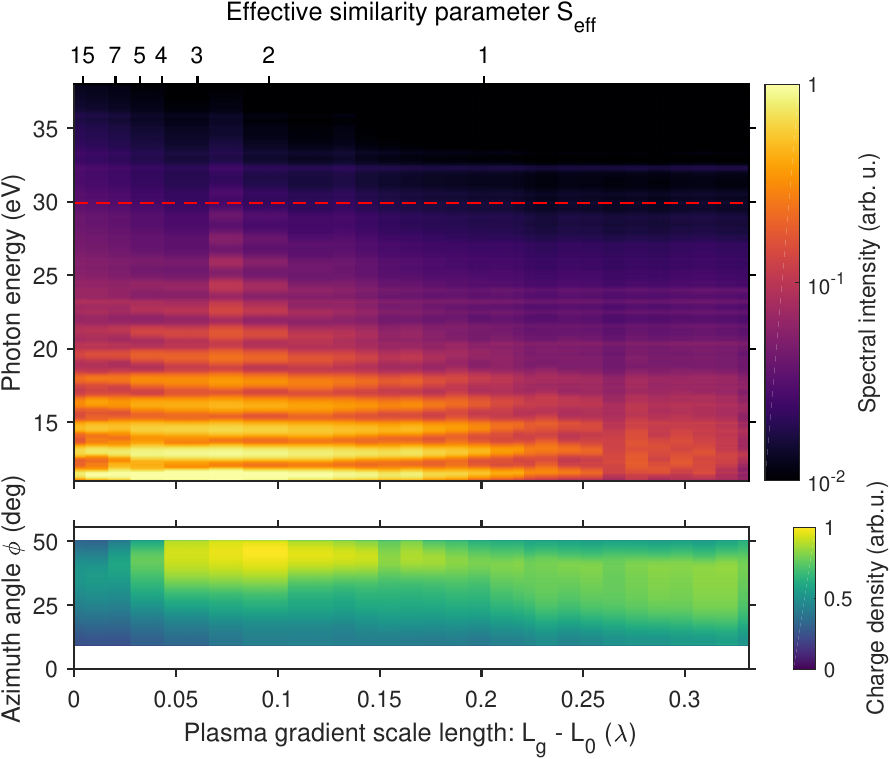}
	\caption{Synchronously measured SHHG (upper, log-scale) and electron (lower) emission as a function of the plasma gradient scale length for an 8-fs, $a_0=1.6$ driver pulse. The dashed red line marks the CWE cutoff energy. The normal and specular directions correspond to $\phi=0$ and $\phi=55^\circ$. The upper horizontal axis gives $S_\mathrm{eff}$ corresponding to $\L$ as given by eq.~\ref{eq:Seff} and assuming $L_0=\lambda/100$.}
	\label{fig:scan9fs}
\end{figure}

Reducing the driving pulse duration to 8~fs at constant pulse energy increases the peak normalized vector potential to $a_0=1.5$ and leads to the results shown in Figure~\ref{fig:scan9fs}. 
As for the 24-fs driver, optimal relativistic SHHG is observed around \commSDI{$\L-L_0\approx\lambda/10$}, with a clear correlation with electron emission for softer gradients. Additionally, we now observe the spectral extent of the SHHG signal to reach and possibly slightly surpass the CWE-cutoff. For longer $\L$, the relativistic SHHG signal persists until \commSDI{$\L-L_0\approx\lambda/4$} but its intensity and spectral extent gradually decreases.

An even shorter 4-fs driving pulse is expected to strongly boost the SHHG signal beyond the CWE cutoff. Self-steepening in the helium-filled hollow-core fiber used for post-compression however reduces the central wavelength to $\lambda\approx770\:$nm so that the peak normalized vector potential only increases to $a_0=2.1$ at constant pulse energy. Furthermore, such short pulses become extremely sensitive to spatio-temporal couplings which can drastically reduce the achievable peak intensity. On many days we thus observe reduced spectral extent of the SHHG spectra generated with $\lesssim4$-fs pulses as compared to those obtained with 8-fs pulses shown in Fig.~\ref{fig:scan9fs}, as was the case for the experiments reported in ref.~\cite{bohle_generation_2020}.

\begin{figure}[tb]
	\centering
	\includegraphics[width=.5\linewidth]{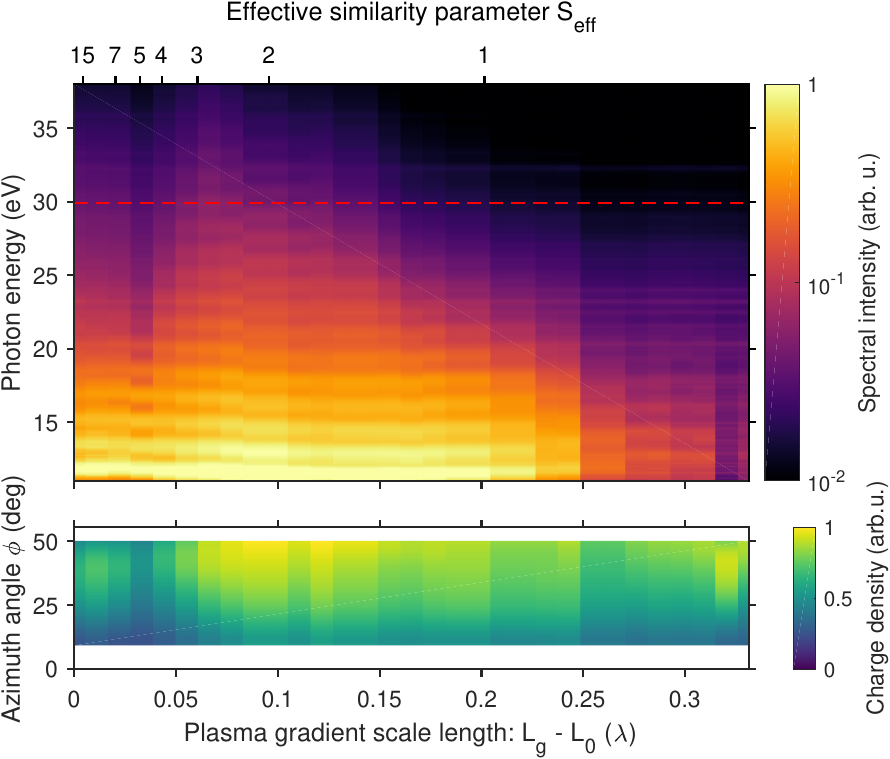}
	\caption{Synchronously measured SHHG (upper, log-scale) and electron (lower) emission as a function of the plasma gradient scale length for a 4-fs, $a_0=2.1$ driver pulse with $\lambda=770\:$nm central wavelength. The dashed red line marks the CWE cutoff energy. The normal and specular directions correspond to $\phi=0$ and $\phi=55^\circ$. The upper horizontal axis gives $S_\mathrm{eff}$ corresponding to $\L$ as given by eq.~\ref{eq:Seff} and assuming $L_0=\lambda/100$.}
	\label{fig:scan4fs}
\end{figure}

We did nonetheless succeed in recording datasets like that shown in Figure~\ref{fig:scan4fs} with a 4-fs driving pulse and shot-to-shot randomly varying CEP. Similarly to the above results, relativistic SHHG is \commSDI{observed} for plasma density gradient scale lengths between \commSDI{$\L-L_0\approx\lambda/20$ and $\lambda/4$}, where now $\lambda=770\:$nm. Around the optimal \commSDI{$\L-L_0\approx\lambda/10$}, the spectra now extend well beyond the CWE cutoff, which is definite proof for relativistic SHHG, reported here for the first time with a 1-kHz repetition rate. The spectra still present a discernible harmonic modulation, in particular for the softer gradients, and one can make out small gradual $\L$-dependent shifts of the harmonic peaks. These may result from a varying temporal structure of the generated very short (essentially 2-pulse~\cite{bohle_generation_2020}) attosecond-pulse-train. All recorded spectra are continous, i.e. the harmonic peaks spectrally overlap, but their averaging over of random driver CEPs precludes us from inferring a temporal structure from spectral features~\cite{kormin_spectral_2018,jahn_towards_2019,bohle_generation_2020}.

The electron emission is again correlated with relativistic SHHG. Compared to figures~\ref{fig:scan24fs} and \ref{fig:scan9fs}, we can make out that the electron beam is less clearly pushed away from the 55$^\circ$-specular direction, because of less efficient ponderomotive scattering of the electrons by the sub-2-cycle laser pulse as compared to the multi-cycle 8-fs and 24-fs pulses.


\section{Discussion}

We have found that relativistic SHHG occurs for plasma density gradient scale lengths \commSDI{$\L\in[\lambda/20,\lambda/4]$, with an optimum at $\L\approx\lambda/10$}, with an assumed laser-contrast induced $L_0\approx\lambda/100$. This optimum plasma gradient scale length is consistent with various earlier experiments with $\approx30$-fs laser drivers focused to higher $a_0>3$~\cite{Roedel2012ultrasteep,kahaly_direct_2013,chopineau_identification_2019,gao_double_2019}. Here we find, for three different driving laser durations from 24 to 4~fs and lower peak intensities corresponding to $a_0=1.1$ to $2.1$, the same optimal $\L$-range. This apparent universality of the optimal plasma density gradient scale length is an intriguing observation.

In order to study density dependencies, particle-in-cell simulations often treat initially step-like plasma density profiles with variable peak density $n$. This adjusts the similarity parameter $S=n/(n_c a_0)$, which describes the relative strengths of the laser and induced internal plasma fields and scales laser plasma-interaction in the relativistic regime~\cite{gordienko_scalings_2005}. Optimal conditions for the relativistic push-pull process, relying on the formation and dynamics of a sharp electron density spike, are predicted for $S\in[1,5]$~\cite{gonoskov_ultrarelativistic_2011,gonoskov_theory_2018,edwards_x-ray_2020}. 

For currently explored experimental laser intensities, $a_0\sim1$, few-times overcritical plasma densities should thus be optimal. In solid-target experiments however, the plasma rather has $\sim100 n_c$ bulk density, and the interaction rather occurs in the smooth density gradient at the surface. In the push-phase, the laser pulse sweeps electrons forward and penetrates this gradient up to a depth $x_\mathrm{b}$, where the laser's Lorentz force and the electrostatic restoring force exerted by the plasma ion background are balanced~\cite{gonoskov_ultrarelativistic_2011, gonoskov_theory_2018}. The plasma ion density at the depth $x_\mathrm{b}$ can then be taken to define an effective similarity parameter $S_\mathrm{eff}$. This was demonstrated in ref.~\cite{blackburn_relativistically_2018} with numerical simulations for a linear density ramp, with the core result that properties of SHHG obtained for a given $S_\mathrm{eff}$ match those obtained with a step-like plasma profile whose density satisfies $S = S_\mathrm{eff}$.

Here, we take the same approach but consider instead the experimentally relevant exponential density profile of eq.~\ref{eq:densityprofile}. 
In this plasma, the laser-plasma force balance yields an instantaneous penetration depth~\cite{vincenti_optical_2014, thevenet_physics_2016} $x_\mathrm{b} = L \ln\left[ a (1+\sin\theta) \cos\theta/ (\pi L/\lambda) \right]$, where $\theta$ is the laser incidence angle on the plasma surface, and $a=a_0\cos(\varphi)$ is the instantaneous value of the normalized vector potential during the pushing half-cycle, i.e. $\varphi\in[-\pi/2,\pi/2]$. \commSDI{With the half-cycle averaged vector potential $(2/\pi)a_0$, we can define an effective penetration depth of 
\begin{equation}
     \overline{x_\mathrm{b}} = \L \ln\left[ \frac{2a_0 (1+\sin\theta) \cos\theta}{\pi^2 (\L/\lambda)} \right].
     \label{eq:xbeff}
\end{equation}
}
Experimentally varying the scale length $\L$ thus corresponds to adjusting the effective plasma (ion) density $n[\overline{x_\mathrm{b}}(\L)]$, where a larger $\L$ corresponds to a lower density. With this density, we find an effective similarity parameter
\commSDI{
\begin{equation}
    S_\mathrm{eff} = \frac{n(\overline{x_\mathrm{b}})}{n_c a_0} = \frac{2(1+\sin\theta) \cos\theta}{\pi^2 (\L/\lambda)}.
    \label{eq:Seff}
\end{equation}
}
It is remarkable that this is \emph{in}dependent of the laser intensity and, for a given incidence angle, only depends on the gradient scale length. For a given exponential density profile, and for any intensity that is sufficiently high to be considered relativistic, the laser thus piles up the electrons at the depth where the ion density corresponds to the \emph{same} similarity parameter. This direct correspondence of $\L$ and $S_\mathrm{eff}$ is the reason for the intensity-independent optimal $L$-range that we find in our experiments. 

\begin{figure}[tb]
	\centering
	\includegraphics[width=.5\linewidth]{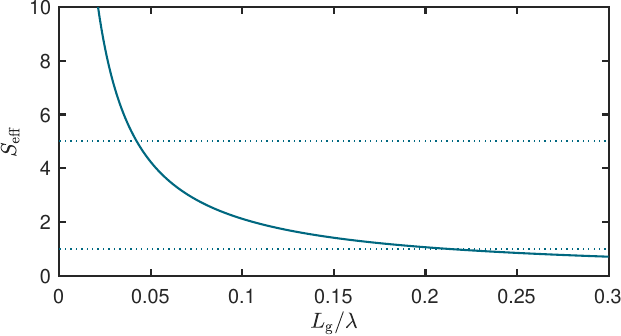}
	\caption{\comm{Theoretical intensity-\emph{in}dependent effective similarity parameter $S_\mathrm{eff}$ (eq.~\ref{eq:Seff})} for $\theta=55^\circ$ as a function of the scale length $\L$ of an exponential plasma density gradient (eq.~\ref{eq:densityprofile}). Dashed lines mark $S_\mathrm{eff}=1$ and $S_\mathrm{eff}=5$.}
	\label{fig:Seff}
\end{figure}

We plot $S_\mathrm{eff}$ from eq.~\ref{eq:Seff} for our experimental incidence angle $\theta=55^\circ$ in figure~\ref{fig:Seff}. The theoretically predicted optimal range $S\in[1,5]$ for relativistic SHHG~\cite{gonoskov_ultrarelativistic_2011,gonoskov_theory_2018,edwards_x-ray_2020} is thus achieved for \commSDI{$\L\in[\lambda/25, \lambda/5]$}. This is in excellent agreement with our experimentally observed intensity-\emph{in}dependent $\L$-range. Assuming $L_0=\lambda/100$, eq.\ref{eq:Seff} yields the $S_\mathrm{eff}$-axes in figures~\ref{fig:scan24fs}--\ref{fig:scan4fs}.

\section{Conclusion}
We have reported clear evidence for the first observation of relativistic SHHG from plasma mirrors driven at kilohertz repetition rate. These are correlated to the emission of energetic electrons through the variation of the surface plasma density gradient scale length $\L$. For three increasingly short and intense driving pulses (from 24~fs (9 optical cycles) and $a_0=1.1$ to 4~fs (1.7 optical cycles) and $a_0\approx2.1$), relativistic SHHG has been observed for the same $\L$-range, \commSDI{$\L\in[\lambda/20,\lambda/4]$}, with an optimum gradient scale length of \commSDI{$\L\approx\lambda/10$}. The universality of the optimal $\L$-range has been rationalized by deriving a direct intensity-\emph{in}dependent link between the scale length $\L$ of an exponential plasma density profile and an effective similarity parameter for relativistic laser-plasma interactions. The observed optimal $\L$-range corresponds very well to theoretically expected optimal similarity parameters. 

These results open the route to a detailed examination and optimization of relativistic SHHG, exploiting the high kHz-repetition rate and exceptional stability of our laser system~\cite{ouille_relativistic-intensity_2020} and next-generation systems in development~\cite{toth_sylos_2020}. In particular for the shortest pulse durations and with added CEP-locking, we should become able to harness the favorable phase properties of relativistic SHHG for the generation of powerful isolated attosecond pulses from ultra-intense laser-driven plasma mirrors~\cite{kormin_spectral_2018,jahn_towards_2019,bohle_generation_2020}.

\medskip

\noindent\textbf{Author Contributions.} R.L.-M., J.F., and S.H. designed the research project. M.B., F.B., and S.H. built the SHHG setup, made preliminary experiments and developed the data analysis routines. M.O., J.K. and A.V. operated the laser chain. M.O., J.K., L.D., D.L. and  S.H. performed the experiments. S.H., M.O. and J.K. analyzed the data. S.H. wrote the paper, with support through comments from all authors.

\medskip

\noindent\textbf{Funding.} This work was supported by the Agence Nationale pour la Recherche (ANR-11-EQPX-005-ATTOLAB, ANR-14-CE32-0011-03 APERO); Investissements d’Avenir program LabEx PALM (ANR-10-LABX-0039-PALM); European Research Council (ERC FEMTOELEC 306708, ERC ExCoMet 694596); LASERLAB-EUROPE (H2020-EU.1.4.1.2. grant agreement ID 654148), R\'egion Ile-de-France (SESAME 2012-ATTOLITE).

\bibliography{BIB_kHzROM}

\end{document}